\pdfoutput=1
\documentclass[11pt,twoside,a4paper,cmspaper,final,collab]{cms-tdr}
\def\svnVersion{57bf0b0}\def\svnDate{2024/03/19}\def\cmsCernNoTag{CERN-EP-2023-203}\def\cmsCernDate{\today}\def\cmsMessage{Published in Physical Review Letters as \href{http://dx.doi.org/10.1103/PhysRevLett.132.121901}{\doi{10.1103/PhysRevLett.132.121901}.}}
\begin{document}\cmsNoteHeader{SMP-22-006}

\newlength\cmsFigWidth
\newlength\cmsTabSkip\setlength{\cmsTabSkip}{1.7ex}
\ifthenelse{\boolean{cms@external}}{\setlength\cmsFigWidth{0.85\columnwidth}}{\setlength\cmsFigWidth{0.4\textwidth}}
\ifthenelse{\boolean{cms@external}}{\providecommand{\cmsTable}[1]{#1}}{\providecommand{\cmsTable}[1]{\resizebox{\textwidth}{!}{#1}}}
\newcommand{\mll}{m_{\Pell\Pell}}
\newcommand{\mjj}{m_\mathrm{jj}}
\newcommand{\mzg}{m_{\PZ\PGg}}
\newcommand{\mllg}{m_{\Pell\Pell\PGg}}
\newcommand{\etajj}{\abs{\Delta \eta_{\mathrm{jj}}}}
\newcommand{\gbarrel}{\PGg_{\text{barrel}}}
\newcommand{\gendcap}{\PGg_{\text{endcap}}}
\newcommand{\dphizgjj}{\Delta \phi_{\PZ\PGg,\mathrm{jj}}}
\newcommand{\mTH}{m_{\mathrm{T}}^{\PH}}
\newcommand{\mTWW}{m_{\mathrm{T}}^{\PW\PW}}
\newcommand{\ptvecll}{\ptvec^{\Pell\Pell}}
\newcommand{\ptll}{\pt^{\Pell\Pell}}
\newcommand{\dphill}{\Delta\phi_{\Pell\Pell}}
\newcommand{\drll}{\Delta R_{\Pell\Pell}}
\newcommand{\ptltwo}{\pt^{\Pell_{2}}}
\newcommand{\ptvecltwo}{\ptvec^{\Pell_{2}}}
\newcommand{\ttst}{\ttbar\PGg+\PQt\PW}
\newcommand{\tautau}{\PGtp{}\PGtm}
\newcommand{\DeepJet}{\ensuremath{\textsc{DeepJet}}\xspace}

\cmsNoteHeader{SMP-22-006}

\title{Observation of \texorpdfstring{$\PW\PW\PGg$}{WW gamma} Production and Search for \texorpdfstring{$\PH\PGg$}{H gamma} Production in Proton-Proton Collisions at \texorpdfstring{$\sqrt{s}=13\TeV$}{sqrt(s) = 13 TeV}}

\date{\today}

\abstract{The observation of $\PW\PW\PGg$ production in proton-proton collisions at a center-of-mass energy of 13\TeV with an integrated luminosity of 138\fbinv is presented. The observed (expected) significance is 5.6 (5.1) standard deviations.  Events are selected by requiring exactly two leptons (one electron and one muon) of opposite charge, moderate missing transverse momentum, and a photon. The measured fiducial cross section for $\PW\PW\PGg$ is $5.9 \pm 0.8 \stat \pm 0.8 \syst \pm 0.7\,\text{(modeling)}\unit{fb}$, in agreement with the next-to-leading order quantum chromodynamics prediction. The analysis is extended with a search for the associated production of the Higgs boson and a photon, which is generated by a coupling of the Higgs boson to light quarks. The result is used to constrain the Higgs boson couplings to light quarks.
}

\hypersetup{
pdfauthor={CMS Collaboration},
pdftitle={Observation of WWgamma production and search for Hgamma production in proton-proton collisions at 13 TeV},
pdfsubject={CMS},
pdfkeywords={CMS,  multi-boson, Yukawa couplings}}

\maketitle

Measurements of multiple electroweak (EW) bosons produced at a common interaction vertex are a key to understanding the EW sector of the standard model (SM). The non-Abelian structure of the EW interaction predicts the presence of self-interactions among the vector bosons ($\PW,\PZ,\PGg$), leading to a rich variety of multiboson production mechanisms. Many multiboson processes are currently accessible only at the CERN LHC given the energies and integrated luminosities required to observe them. The CMS and ATLAS Collaborations have both recently observed the simultaneous production of three massive gauge bosons~\cite{CMS:2020hjs,ATLAS:2021atz}. Additionally, ATLAS has reported the observation of $\PW\PZ\Pgg$~\cite{ATLAS:WZg} with an observed significance of 6.3 standard deviations. The double-photon production processes $\PW\Pgg\Pgg$ and $\PZ\Pgg\Pgg$ have also been measured by both CMS and ATLAS~\cite{ATLAS:Wgg,ATLAS:Zgg,CMS:2021jji}. Searches for $\PW\PW\PGg$ production have previously been conducted by both CMS and ATLAS at a center-of-mass energy of 8\TeV~\cite{CMS:2014cdf,ATLAS:2017bon} where only upper limits were set because of a lack of statistical power and sensitivity.

Triboson production includes not only the interactions involving triple and quartic gauge couplings (TGCs and QGCs), but also the mediation of the Higgs boson ($\PH$), providing an opportunity to measure or constrain Yukawa couplings. Deviations from theoretical predictions in the triboson measurements could provide indirect evidence of new particles or new interactions. Recently, proposals to exploit $\PH\PGg$ production to probe Higgs boson couplings with light (\PQc, \PQs, \PQu, and \PQd) quarks have been published~\cite{Khanpour:2017inb,PhysRevD.103.095023,Falkowski:2020znk}. Since the gluon-initiated contribution $\Pg\Pg\to \PH\PGg$ vanishes according to Furry's theorem~\cite{Furry,Khanpour:2017inb}, $\PH\PGg$ inclusive production at the LHC is directly related to the Higgs boson Yukawa couplings to the light quarks. Various interpretations~\cite{Zhou:2665087,PhysRevD.105.096027,PhysRevD.93.013019} of the light quark Yukawa couplings were previously proposed. Similarly, gluon fusion production can constrain the light quark Yukawa couplings~\cite{deFlorian:2016spz}. Recently, CMS reported a direct constraint on the charm quark Yukawa coupling modifier of $1.1<\abs{\kappa_{\PQc}}<5.5$ at 95\% confidence level (C.L.)~\cite{CMS:2021zwx} and ATLAS provided an upper bound of $\abs{\kappa_{\PQc}}<8.5$ at 95\% C.L.~\cite{ATLAS:2021zwx}. However, the upper bounds on the strange quark Yukawa coupling are presently significantly less stringent~\cite{PhysRevD.101.115005,Aaboud2018}.

\begin{figure*}[htb!]
    \centering{
    \includegraphics[width=0.24\textwidth]{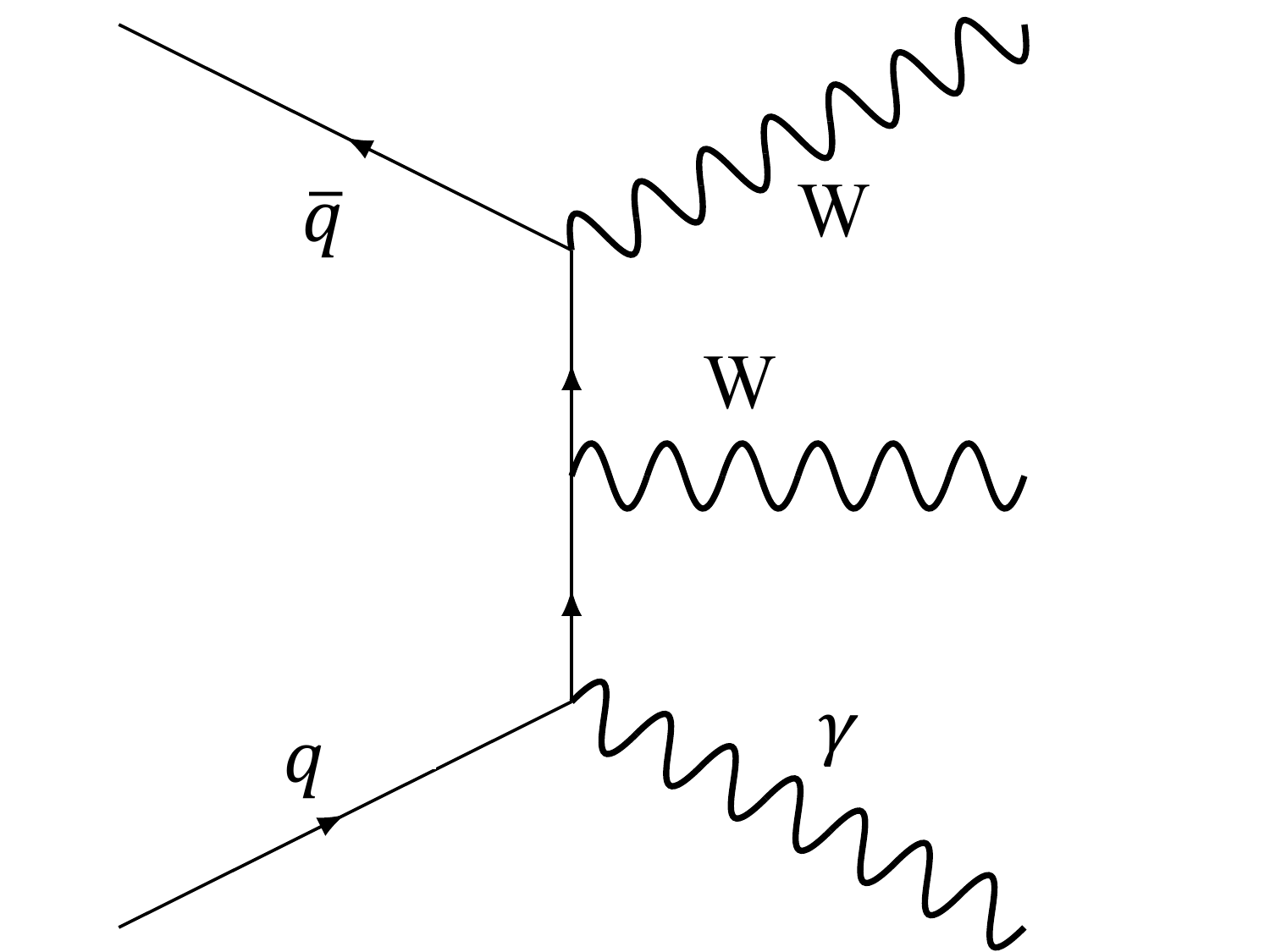}
    \includegraphics[width=0.24\textwidth]{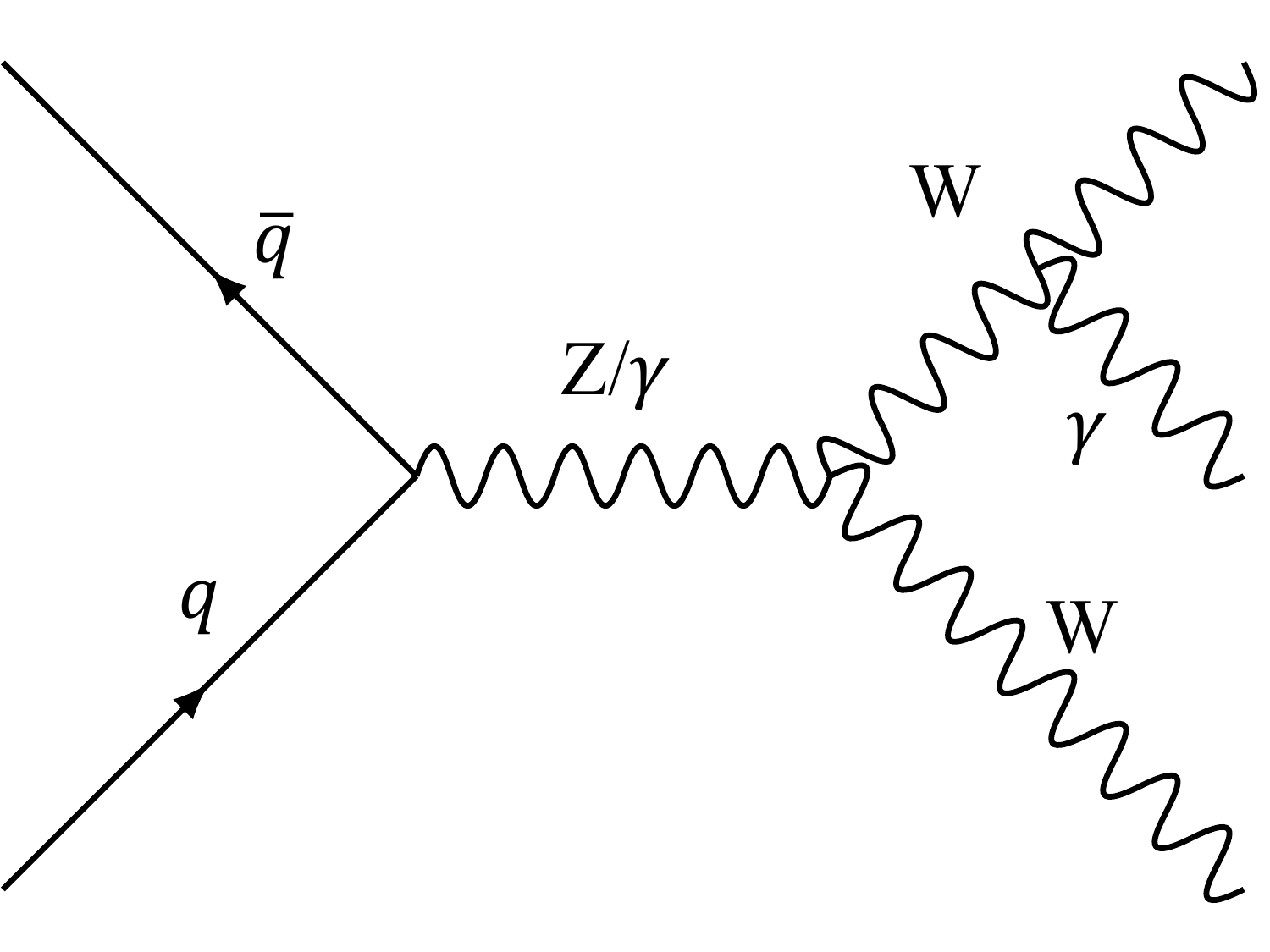}
    \includegraphics[width=0.24\textwidth]{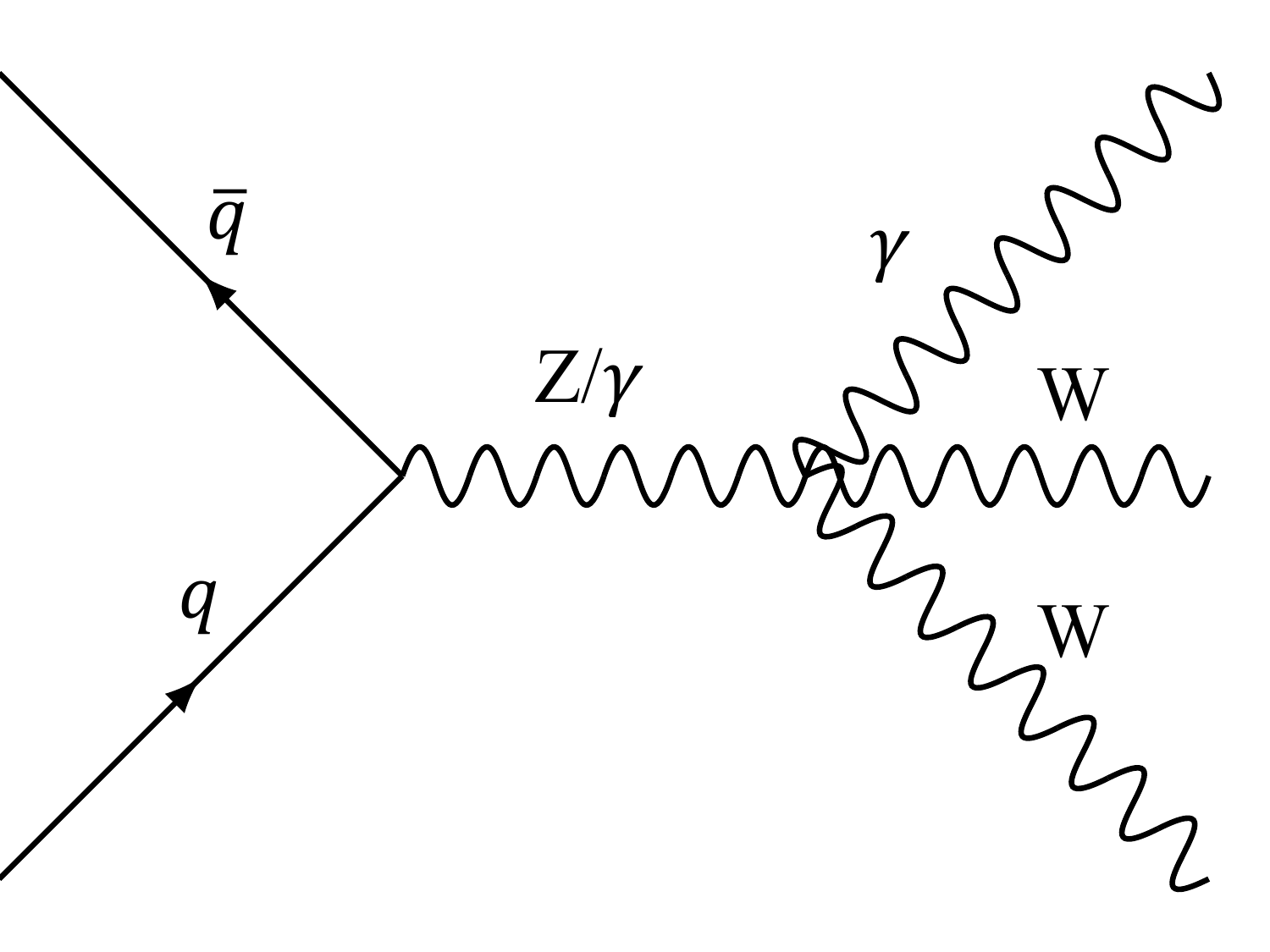}
    \includegraphics[width=0.24\textwidth]{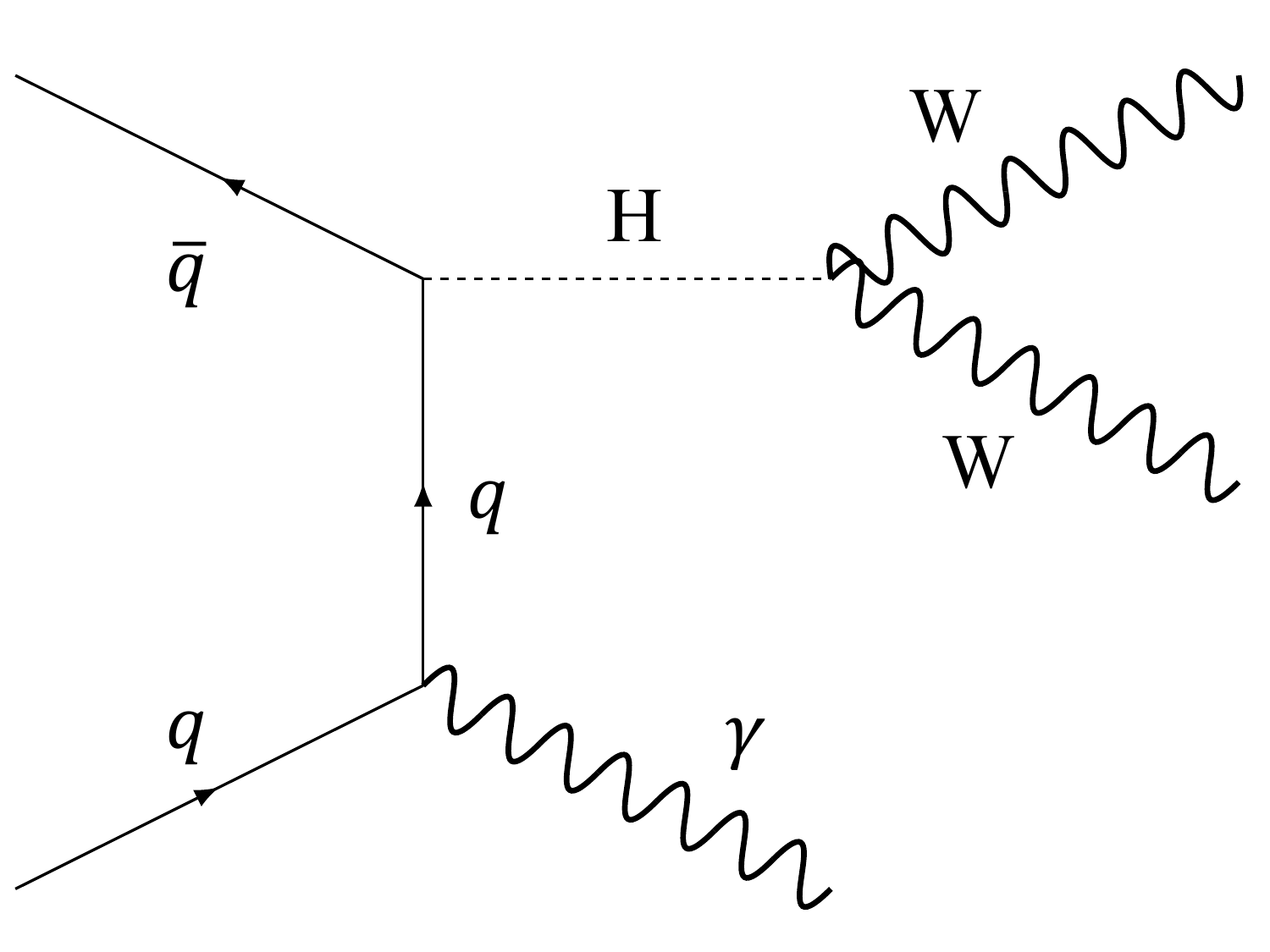}
    }
    \caption{Representative Feynman diagrams for the $\PW\PW\PGg$ process at LO, from left to right: ISR, TGC, QGC, and $\PH\PGg$ associated production.}
    \label{fig:LO diagrams}
\end{figure*}

The EW $\Pep\PGne\PGmm\PAGnGm\Pgg$ and $\PGmp\PGnGm\Pem\PAGne\Pgg$ production in proton-proton ($\Pp\Pp$) collisions at leading order (LO) can proceed via (i) initial-state radiation (ISR) from one of the incoming quarks; (ii) final-state radiation (FSR) from the outgoing charged leptons; (iii) the $\PW\PW\PZ$ or $\PW\PW\Pgg$ TGC; (iv) the $\PW\PW\PZ\Pgg$ or $\PW\PW\Pgg\Pgg$ QGC; and (v) the associated production of the Higgs boson and a photon. Figure~\ref{fig:LO diagrams} shows examples of these processes. At higher orders in quantum chromodynamics (QCD)~\cite{Zhu:2020ous}, additional quarks can appear in the final state, and photons can arise via FSR from an outgoing quark or lepton. 

This Letter reports a first observation of $\PW\PW\PGg$ production as well as a search for $\PH\PGg$ production generated through the Higgs boson interactions with light quarks. The measurements are based on $\sqrt{s} = 13\TeV$ $\Pp\Pp$ collision data collected with the CMS detector during 2016--2018, with an integrated luminosity of $138\fbinv$. Tabulated results are provided in HEPData~\cite{HEPData}.

The central feature of the CMS apparatus is a superconducting solenoid of 6\unit{m} internal diameter, providing a magnetic field of 3.8\unit{T}. Within the magnetic field, there are silicon pixel and strip trackers plus several calorimeters, a lead tungstate crystal electromagnetic calorimeter (ECAL) and a brass-scintillator calorimeter; both of these are composed of a barrel and two end cap sections. The forward calorimeters extend the pseudorapidity ($\eta$) coverage provided by the barrel and end cap detectors. The particle-flow (PF) algorithm~\cite{CMS:2017yfk} reconstructs and identifies each individual particle in an event, with an optimized combination of information from the various elements of the CMS detector. Muons are detected in gas-ionization chambers embedded in the steel return yoke outside the solenoid. A detailed description of the CMS detector, together with a definition of the coordinate system used and the relevant kinematic variables, is reported in Ref.~\cite{Chatrchyan:2008zzk}.

Electrons and photons are measured in the range of $\abs{\eta} < 2.5$ corresponding to the acceptance of the tracker. The energy of an electron is obtained from a combination of three measurements: the electron momentum at the primary interaction vertex as determined by the tracker~\cite{pvdefinition}; the energy of the corresponding ECAL cluster; and the energy sum of all bremsstrahlung photons spatially compatible with the original electron track. The photon momentum is determined by using an energy measurement in the ECAL cluster, which is inconsistent with any charged-particle track in the tracker~\cite{cmscollaboration2020electron}. Muons are measured within $\abs{\eta} < 2.4$ and their momenta are determined by using a global fit of muon measurements in the gas-ionization chambers and matched tracks in the silicon tracker~\cite{muon_13tev}.

Jets are reconstructed from PF candidates~\cite{CMS:2017yfk} using the anti-\kt jet clustering algorithm with a distance parameter of 0.4~\cite{antikt,Cacciari:2011ma}. The energy is obtained within $\abs{\eta}<4.7$, from the corresponding corrected energy deposits in the ECAL and hadron calorimeters that are matched to charged hadron tracks in the tracker; it is corrected for: (i) pileup (PU) from multiple interactions per proton bunch crossing in the colliding beams; (ii) nonuniformity of the detector response; and (iii) residual differences between data and simulation. The average neutral energy density from PU is estimated and subtracted from the reconstructed jet energies and the energy sum used in the calculation of lepton isolation~\cite{CACCIARI2008119}. Jets containing the decay of $\PQb$ quarks (b jets) are identified using the medium working point of the \DeepJet algorithm~\cite{Bols:2020bkb} with a misidentification probability of 1\%.

The missing transverse momentum vector ($\ptvecmiss$) is computed as the negative vector transverse momentum (\ptvec) sum of all PF candidates in an event~\cite{CMS:2017yfk} and its magnitude is denoted by $\ptmiss$~\cite{Sirunyan_2019}. The $\ptvecmiss$ of an event is intended to represent the neutrinos associated with a single $\Pp \Pp$ interaction within a bunch crossing. The $\ptvecmiss$ is also modified to include corrections to the energy scale and resolution of the reconstructed jets in the event.

The $\PW\PW\PGg$ signal is simulated at next-to-LO (NLO) in QCD and includes intermediate $\tau$ decays. The $\PH\PGg$ signal is simulated at NLO in QCD for $\PQc\PAQc$ production and at LO using the Higgs effective Lagrangian model~\cite{alloul2014phenomenology} for other light quarks where a generator-level cut of $\pt^\PGg>5\GeV$ is applied. The initial values of the quark masses are taken from the Particle Data Group~\cite{quarkmass} and a running coupling for the $\PQc$ quark is implemented. Both these simulations use the \MGvATNLO v2.6.5~\cite{MGatNLO} Monte Carlo event generator, and the Higgs boson decay modeled by the \textsc{JHUGen} v7.2.7 for the $\PH\PGg$ signal~\cite{Gao:2010qx,Bolognesi:2012mm,Anderson:2013afp,Gritsan:2016hjl}. The parton showering and hadronization are performed using {\PYTHIA} 8.226~\cite{Sjostrand:2014zea}, and the detector simulation is performed using \GEANTfour v10.4.3~\cite{geant4}. The {\PYTHIA}8 CP5 underlying event tune~\cite{Sirunyan:2019dfx} with NNPDF3.1 next-to-NLO parton distribution functions (PDFs) is used. Neither EW nor next-to-NLO QCD corrections are applied.

Background processes containing prompt leptons and a prompt photon, including $\PZ\Pgg$ production, $\ttbar\Pgg$ production, and associated production of a single top quark and $\PW$ boson, are simulated using \MGvATNLO or \POWHEG~v2.0~\cite{Nason:2004rx, Frixione:2007vw, Alioli:2010xd, Alioli:2010xa,Alioli:2008tz, Bagnaschi:2011tu} at NLO in QCD interfaced with \PYTHIA{8} for hadronization and fragmentation in a manner similar to that for the $\PW\PW\PGg$ signal sample. The background due to events containing nonprompt leptons and photons, including those from instrumental mismeasurements and genuine leptons or photons within jets, is estimated from data using a method similar to that of Ref.~\cite{CMS:2021foa,PhysRevD.108.032017,PhysRevD.104.072001}. The relative contribution of events with well-isolated, high-quality leptons to less-isolated, lower-quality leptons is measured in a dijet control region (CR) in data as a function of the lepton $\abs{\eta}$ and $\pt$, and corrected for prompt leptons and prompt photon conversions based on simulated samples. A similar procedure is applied for photons, based on a $\PW$+jets CR that excludes the signal region (SR). In the nonprompt-photon case, a fit to the width of the photon ECAL shower is used to determine the nonprompt-photon fraction in the well-isolated, high-quality category, as described in Ref.~\cite{Sirunyan_2020}. Based on the matching to the generator level, the two procedures are combined to avoid double counting~\cite{CMS:2021foa}. The SM contributions from other Higgs-related processes~\cite{Gabrielli2016} are negligible. 

Experimentally, we select $\PW^+\PW^-\Pgg\to$ $\Pep\PGne\PGmm\PAGnGm\Pgg$ and $\PGmp\PGnGm\Pem\PAGne\Pgg$ events, which pass the level-1~\cite{Sirunyan:2020zal} and high-level~\cite{trigger_system} triggers that require an isolated muon and/or electron. We require the isolated electron and muon to satisfy additional identification criteria~\cite{muon_13tev,cmscollaboration2020electron}, a single reconstructed photon~\cite{cmscollaboration2020electron} must be present in the event, and the $\ptmiss$ must exceed 20\GeV. The photon must satisfy high performance identification requirements that correspond to a signal efficiency $>$ 80\%~\cite{cmscollaboration2020electron}. Off-line kinematic requirements on the selected objects, based on the detector acceptance and the trigger thresholds, are $\pt^\PGg>20\GeV$,  $\abs{\eta^\PGg}<2.5$,  $\abs{\eta^{\Pe(\PGm)}}< 2.5$ (2.4) and $\pt^{\Pe(\PGm)} > $ 25 (20)\GeV. To reduce backgrounds from $\PW\PZ\Pgg$ and relevant top quark processes, events are rejected that contain at least one \PQb jet or an additional muon or electron with $\pt > 10\GeV$ passing looser criteria than those of the primary leptons. Moreover, it is required that $\Delta R = \sqrt{\smash[b]{(\Delta\eta)^2 + (\Delta\phi)^2}} > 0.5$, where $\Delta \phi$ and $\Delta \eta$ are the spatial separations in the azimuthal angle $\phi$ and $\eta$ between leptons and photon. We further suppress background contributions by requiring the dilepton mass ($\mll$) $>10$\GeV, the transverse momentum ($\ptll$) $>15$\GeV, and the transverse mass, $\mTWW = \smash[b]{\sqrt{2 \ptll \ptmiss [1 - \cos\Delta\phi(\ptvecll, \ptvecmiss)]}} >10\GeV$. 

A CR with charged leptons of the same sign, SS$\PW\PW\PGg$, is constructed to validate the nonprompt lepton background modeling. Another Top$\PGg$ CR, dominated by events corresponding to top quark production, is used to validate the modeling of both nonprompt-lepton and nonprompt-photon backgrounds. These two CRs are included in the simultaneous maximum likelihood fit to constrain the estimates of these process rates. The selection for the SS$\PW\PW\PGg$ CR is the same as for the SR, except that the $\mTWW$ requirement is removed and the two leptons are required to have the same sign. The definition of the Top$\PGg$ CR also follows closely that of the SR, except that
at least one {\PQb}-tagged jet with $\pt > 20\GeV$ is required and the $\mTWW$ requirement is removed. 

\begin{figure*}[htp]
     \centering      
      \includegraphics[width=0.48\textwidth]{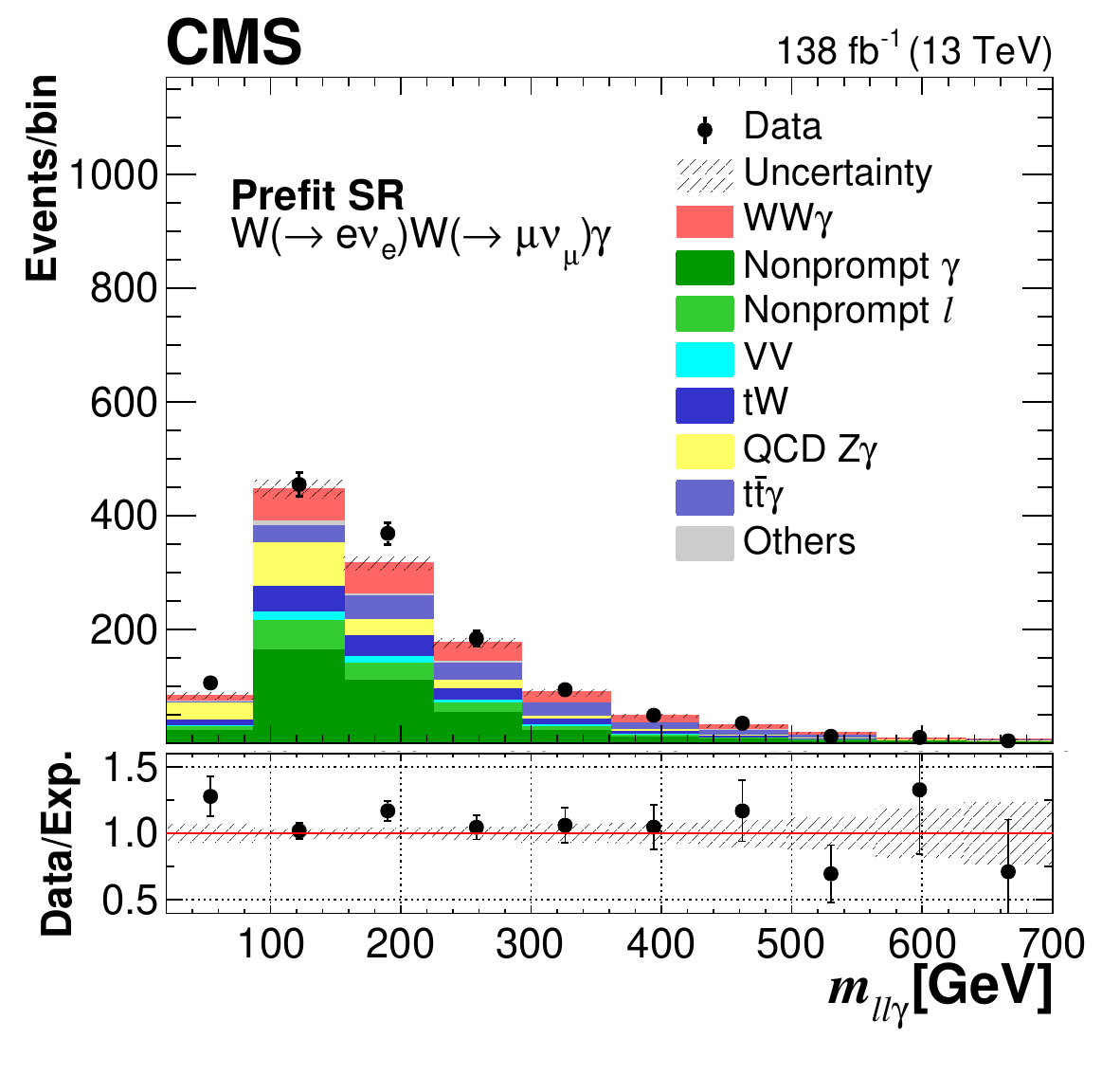}
      \includegraphics[width=0.48\textwidth]{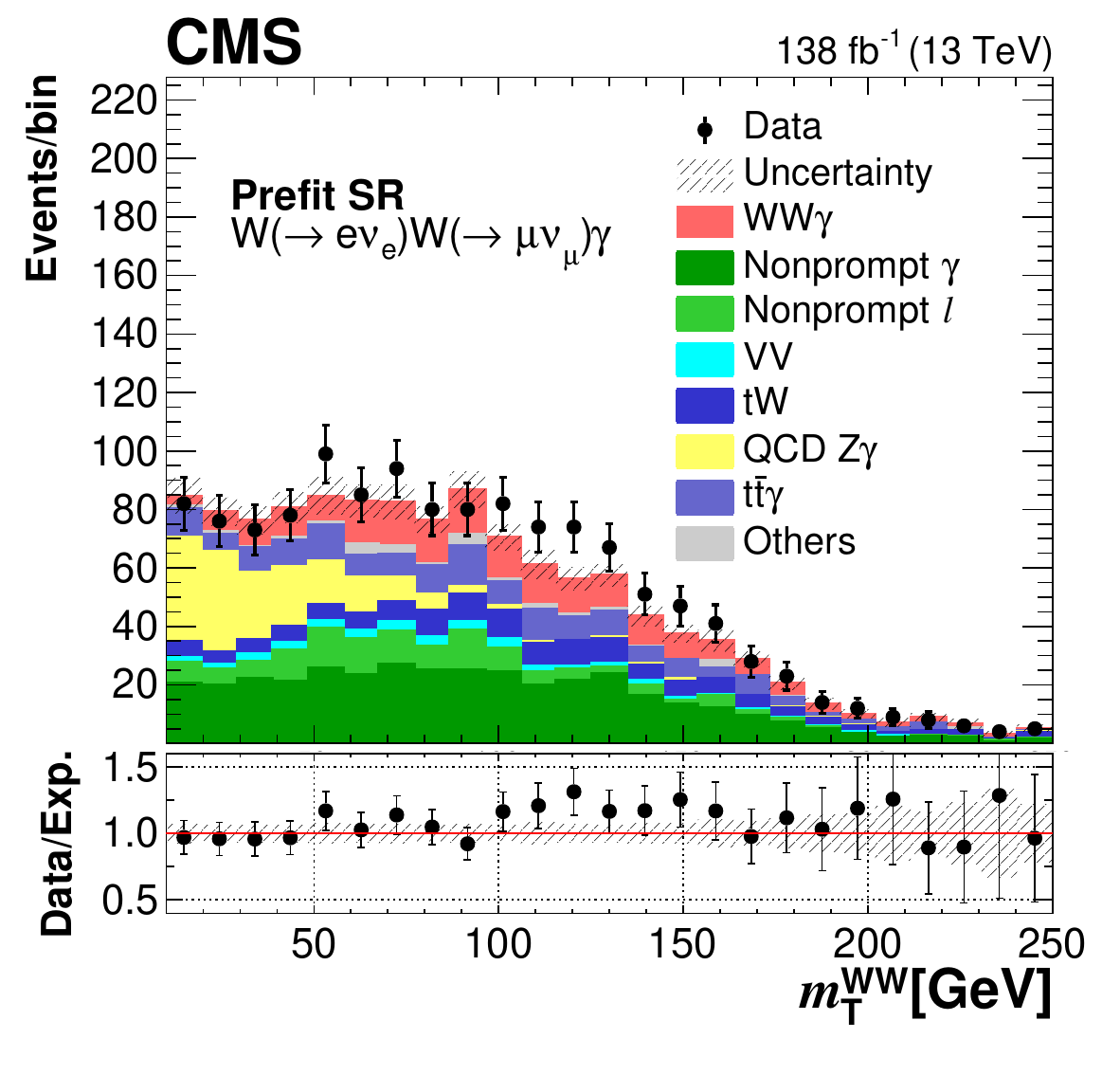}      
    \caption{The distributions of the invariant mass of the dilepton-photon system (left) and the transverse mass of the $\PW\PW$ system (right) after the signal region selection and before the fit to the data. The black points with error bars represent the data and their statistical uncertainties, whereas the shaded band represents the Monte Carlo statistical uncertainties.}
        \label{fig:plots_WWgammarun2}
     
\end{figure*}

The observed distributions in the SR of the invariant mass of the dilepton-photon system ($\mllg$) and $\mTWW$ are compared with the expected distributions before the fit in Fig.~\ref{fig:plots_WWgammarun2}. The experimental data agree with the prediction within the uncertainties. 

Various sources of systematic uncertainty are included in the fit as nuisance parameters and subject to log-normal constraints. Theoretical sources of systematic uncertainty include the choice of the renormalization and factorization scales, PDFs, and parton shower modeling. The two scales are varied by factors of 2 and 0.5 independently. The envelope of these variations, excluding the two extreme (2, 0.5) and (0.5, 2) cases, is assumed as the uncertainty. The systematic uncertainty due to PDFs is calculated using the PDF4LHC15\_nnlo\_30\_pdfas PDF replicas, following the PDF4LHC group prescription~\cite{Butterworth_2016,Harland_Lang_2015,nnpdf,PhysRevD.93.033006}. Parton shower modeling uncertainties, which arise from the renormalization of QCD-induced ISR and FSR, are the dominant uncertainty in the measurement. Variations of the contributions of the ISR and FSR consist of 4 combinations made by keeping one constant and doubling or halving the other. Multiple experimental sources of systematic uncertainties are also included. The most significant contribution arises from the method used to estimate the nonprompt background (8\%), followed by the \PQb tagging efficiency (7\%), and jet energy scale and resolution (5\%), which also affect $\ptvecmiss$. The uncertainties associated with the lepton and photon identification efficiencies, the lepton trigger, PU, and integrated luminosity~\cite{CMS-LUM-17-003,CMS:2018elu,CMS:2019jhq} are also assessed. The statistical uncertainties from the limited size of the simulated signal and background samples are also included in the final fit.

\begin{figure*}[htb!]
     \centering      
      \includegraphics[width=0.49\textwidth]{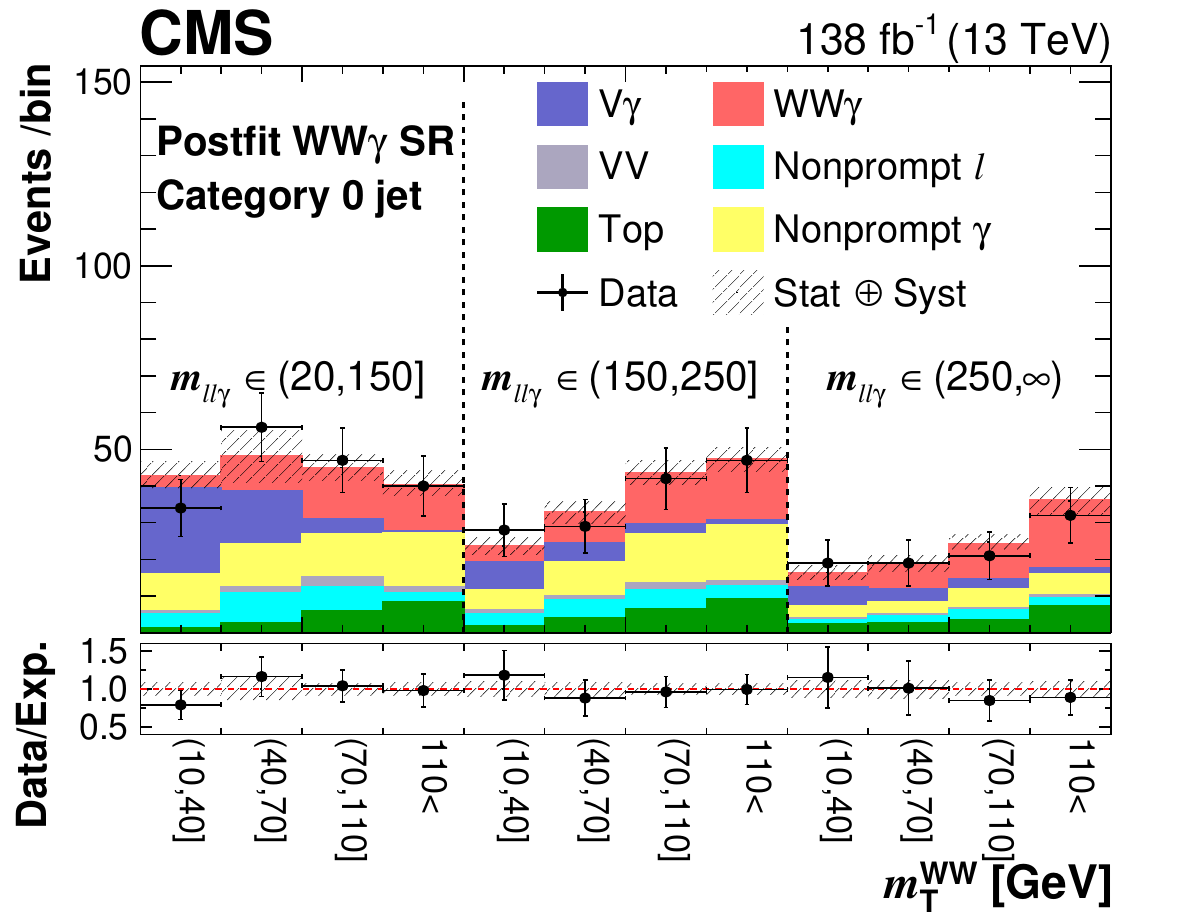}
      \includegraphics[width=0.49\textwidth]{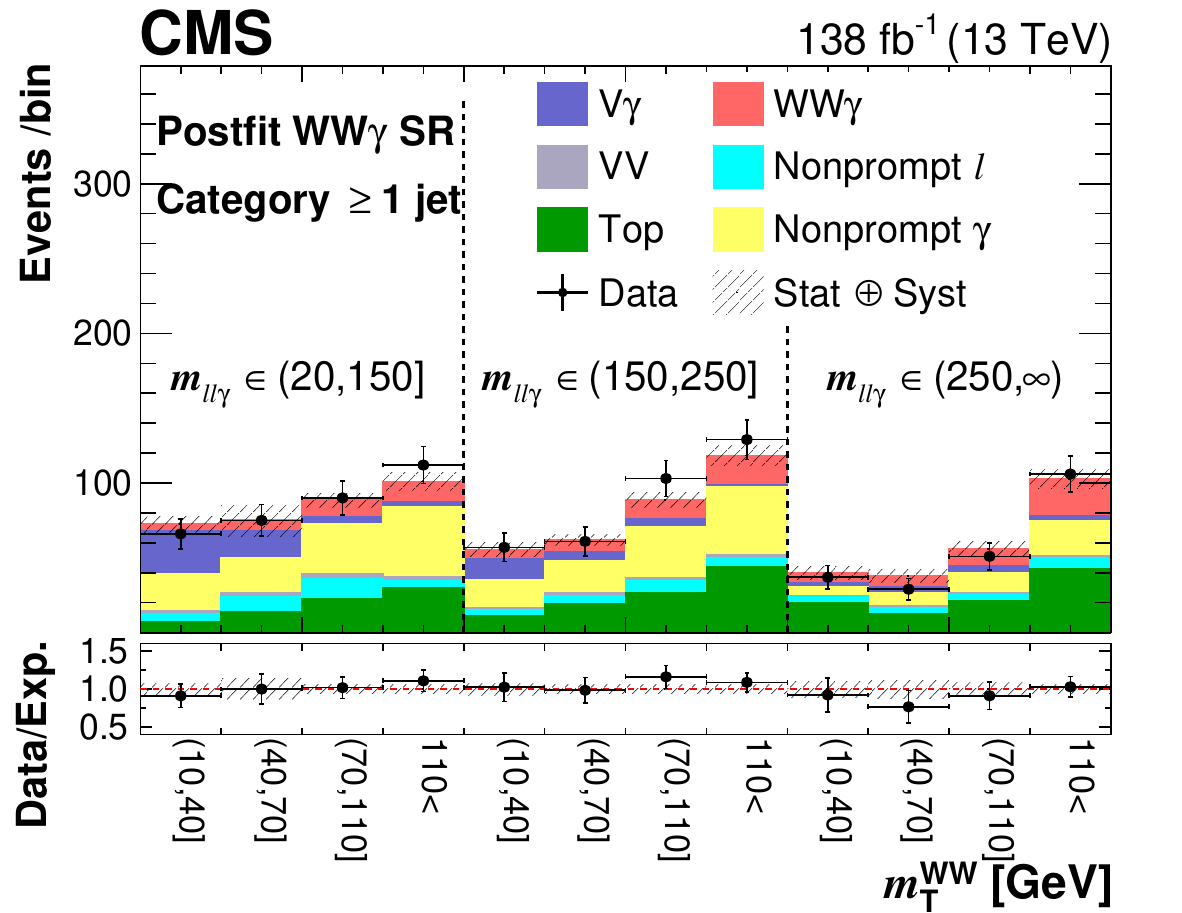}
    \caption{The unrolled two-dimensional $\mTWW$-$\mllg$ distributions in category 0 jet (left) and $\geq$1 jet (right) after the fit to data. The data are compared with the sum of the signal and expected background. The black points with error bars represent the data and their statistical uncertainties, whereas the hatched bands represent the total uncertainties in the predictions.}
        \label{fig:2dfitrun2}
     
\end{figure*}

\begin{table*}[!ht]
\newcolumntype{x}{D{,}{\,\pm\,}{5.3}}
\centering
\caption{The number of events in data and prediction after the fit to data in the SR, SS$\PW\PW\PGg$ CR and Top$\PGg$ CR. The uncertainties include both statistical and systematic contributions.}
\begin{scotch*}{lxxxxx}
Process & \multicolumn{1}{c}{SR (0 jet)} & \multicolumn{1}{c}{SR ($\geq$1 jet)} & \multicolumn{1}{c}{SR (total)} & \multicolumn{1}{c}{SSWW$\PGg$ CR} & \multicolumn{1}{c}{Top$\PGg$ CR} \\
\hline
$\PW\PW\PGg$ & 122,22 & 134,26  & 256,47 & 1.3,0.2  & 12.9,2.7\\[2pt]
QCD $\PV\PGg$ & 72.0,6.0 & 94.6,9.3 & 167,14  & 12.2,2.3  & 12.6,1.2  \\
$\PV\PV$ & 15.1,1.2 & 21.6,2.3 & 36.7,3.5 & 24.9,1.7 & 2.0,0.3 \\
Top & 56.6,6.5 & 271,27  & 328,32  & 2.4,0.6  & 2434,85\\
Nonprompt $\Pell$ & 45.7,4.0 & 77.1,6.2 & 122.9,9.7 & 197,14  & 40,11\\
Nonprompt $\PGg$ & 109.0,8.4 & 300,24 & 410,32 & 19.8,1.6  & 792,62\\
Total  & 420,17 & 898,26 & 1318,43 & 257,14  & 3294,57\\
Data & \multicolumn{1}{c}{414} & \multicolumn{1}{c}{916} & \multicolumn{1}{c}{1330} & \multicolumn{1}{c}{259} & \multicolumn{1}{c}{3287} \\
\end{scotch*}
\label{tab:yields}
\end{table*}
The signal significance and strength are extracted from a binned maximum likelihood fit using two-dimensional distributions in bins of $\mTWW$ and $\mllg$, where the product of the Poisson probability mass functions for each bin forms the likelihood function. The $\Pe\PGm$ final states that result from the intermediate $\tau$ in our signal simulation, are treated as part of the signal in significance and strength extraction. A simultaneous fit including the SR and CRs is performed. Since background processes tend to be concentrated in the $\geq$1 jet region, the SR is divided into two categories based on jet multiplicity: 0 jet and $\geq$1 jet. The number of events in data and predictions after the fit to the data are listed in Table~\ref{tab:yields}. The observed (expected) signal significance from the fit is $5.6\,(5.1)$ standard deviations, corresponding to the observed distributions after the fit to the data shown in Fig.~\ref{fig:2dfitrun2}.  
The observed signal strength, $\mu_{\text{obs.}}=1.11 \pm 0.16 \stat \pm 0.15 \syst \pm 0.13\,\text{(modeling)}$, is extracted in a fiducial region defined by applying the signal selection at particle level, without the requirements on \PQb jets and additional leptons. The theoretical prediction for the $\PW\PW\PGg$ fiducial cross section is $5.33 \pm 0.34\,\text{(scale)} \pm 0.05\,\text{ (PDF)}\unit{fb}$ at NLO QCD as evaluated by \MGvATNLO. The $\PW\PW\PGg$ measured cross section from the simultaneous fit with the uncertainties divided into statistical, experimental, and theoretical modeling components is $\sigma =5.9 \pm 0.8 \stat \pm 0.8 \syst \pm 0.7\,\text{(modeling) }\unit{fb} = 5.9 \pm 1.3 \unit{fb}$. The theoretical modeling uncertainties include the renormalization and factorization of QCD scales, PDFs, and parton shower modeling from all simulations.

We also search for the $\PH\PGg$ production mechanism shown in Fig.~\ref{fig:LO diagrams} with modified Higgs boson couplings to light quarks, which have different $\pt^\PGg$ spectra and equivalently $\PH\PGg$ invariant mass compared with other anomalous $\PH\PZ\PGg$ coupling processes as described in Ref.~\cite{PhysRevD.105.096027}. The selection for this search is similar to the EW $\PW\PW\PGg$ signal selection but targets the Higgs boson characteristics by requiring $\dphill<2.5$, $\drll<2.3$, and $\Delta R_{\Pell\PGg} >0.8$, since the two oppositely charged $\PW$ bosons from the Higgs boson decay tend to have opposite spin orientation and the leptons from $\PW$ bosons are likely to travel in the same direction~\cite{Dittmar:1996ss}. Now the observed $\PW\PW\PGg$ is regarded as a background whose normalization floats and is constrained by incorporating the remaining $\PW\PW\PGg$ events and all CRs in the simultaneous fit. Since the $\Delta R_{\Pell\Pell}$ observable has good discrimination power~\cite{Duehrssen:1261372}, the profile likelihood ratio test statistic~\cite{CMS-NOTE-2011-005} is built separately for four processes in bins of $\Delta R_{\Pell\Pell}$ and $\mTWW$, where $\Delta R_{\Pell\Pell}$ and $\mTWW$ are divided into bins of [0.5, 1.8, 2.0, 2.3) and [0, 10, 40, 70, 110, $\infty$), respectively. The upper limits on the $\PH\PGg$ cross sections at 95\% C.L. are shown in Table~\ref{tab:Limits Hgamma1}. The results can be interpreted as limits on the Higgs boson to light quarks Yukawa couplings $\kappa_{\PQq}$~\cite{PhysRevD.103.095023}, assuming that the light quark and the Higgs boson interaction vertex in Fig.~\ref{fig:LO diagrams} is the only parameter that does not behave according to the SM.
 The normalized light Yukawa couplings $\overline{\kappa}_{\PQq}$ are also provided, which rescales $\kappa_{\PQq}$ into units of $y^{SM}_{\PQb}$ evaluated
at scale $\mu$ = 125\GeV as described in Ref.~\cite{Yu2017}.

\begin{table*}[htb!]
\centering
\caption{Upper limits on the cross section and derived limits in terms of Yukawa coupling at 95\% C.L. for $\PH\PGg$ production initiated by light quarks. }
\cmsTable{
\begin{scotch*}{ccll}
Process  & $\sigma$ upper limits obs. (exp.) [fb]  & $\kappa_{\PQq}$ limits obs. (exp.) at 95\% C.L. &  $\overline{\kappa}_{\PQq}$ limits obs. (exp.) at 95\% C.L.
 \\\hline
$\PQu \PAQu \to \PH+\PGg \to \Pe\PGm\PGn_\Pe\PGnGm\PGg$  & 86 (67)  & $ \lvert\kappa_{\PQu}\rvert  \leq $ 16000 (13000) & $ \lvert\overline{\kappa}_{\PQu}\rvert  \leq $ 7.5 (6.1)\\[2pt]
$\PQd \PAQd \to \PH+\PGg \to \Pe\PGm\PGn_\Pe\PGnGm\PGg$   & 72 (58)  & $ \lvert\kappa_{\PQd}\rvert  \leq $ 17000 (14000) & $ \lvert\overline{\kappa}_{\PQd}\rvert  \leq $ 16.6 (14.7) \\[2pt]
$\PQs \PAQs \to \PH+\PGg \to \Pe\PGm\PGn_\Pe\PGnGm\PGg$   & 66 (49)  & $ \lvert\kappa_{\PQs}\rvert  \leq $ 1700 (1300) & $ \lvert\overline{\kappa}_{\PQs}\rvert  \leq $ 32.8 (25.2)
\\[2pt]
$\PQc \PAQc \to \PH+\PGg \to \Pe\PGm\PGn_\Pe\PGnGm\PGg$   & 88 (66)  & $ \lvert\kappa_{\PQc}\rvert  \leq $ 190 (110) & $ \lvert\overline{\kappa}_{\PQc}\rvert  \leq $ 43.2 (25.0) \\
\end{scotch*}
\label{tab:Limits Hgamma1}}
\end{table*}

In summary, this Letter reports the first observation of $\PW\PW\PGg$ production in proton-proton collisions. The measurement uses a dataset collected by the CMS experiment at the LHC in 2016--2018 at a center-of-mass energy of 13\TeV, with an integrated luminosity of 138\fbinv. The measured fiducial cross section for $\PW\PW\PGg$ production is $5.9 \pm 1.3 \unit{fb}$, in agreement with the prediction at next-to-leading order in quantum chromodynamics. 
A search for the associated production of the Higgs boson and a photon is also performed using the Higgs boson decay to $\PWp\PWm$. A set of limits at 95\% confidence level on the Higgs boson couplings to light quarks is reported.

\begin{acknowledgments}
We congratulate our colleagues in the CERN accelerator departments for the excellent performance of the LHC and thank the technical and administrative staffs at CERN and at other CMS institutes for their contributions to the success of the CMS effort. In addition, we gratefully acknowledge the computing centers and personnel of the Worldwide LHC Computing Grid and other centers for delivering so effectively the computing infrastructure essential to our analyses. Finally, we acknowledge the enduring support for the construction and operation of the LHC, the CMS detector, and the supporting computing infrastructure provided by the following funding agencies: SC (Armenia), BMBWF and FWF (Austria); FNRS and FWO (Belgium); CNPq, CAPES, FAPERJ, FAPERGS, and FAPESP (Brazil); MES and BNSF (Bulgaria); CERN; CAS, MoST, and NSFC (China); MINCIENCIAS (Colombia); MSES and CSF (Croatia); RIF (Cyprus); SENESCYT (Ecuador); MoER, ERC PUT and ERDF (Estonia); Academy of Finland, MEC, and HIP (Finland); CEA and CNRS/IN2P3 (France); SRNSF (Georgia); BMBF, DFG, and HGF (Germany); GSRI (Greece); NKFIH (Hungary); DAE and DST (India); IPM (Iran); SFI (Ireland); INFN (Italy); MSIP and NRF (Republic of Korea); MES (Latvia); LAS (Lithuania); MOE and UM (Malaysia); BUAP, CINVESTAV, CONACYT, LNS, SEP, and UASLP-FAI (Mexico); MOS (Montenegro); MBIE (New Zealand); PAEC (Pakistan); MES and NSC (Poland); FCT (Portugal); MESTD (Serbia); MCIN/AEI and PCTI (Spain); MOSTR (Sri Lanka); Swiss Funding Agencies (Switzerland); MST (Taipei); MHESI and NSTDA (Thailand); TUBITAK and TENMAK (Turkey); NASU (Ukraine); STFC (United Kingdom); DOE and NSF (USA).
\end{acknowledgments}

\bibliography{auto_generated}
\cleardoublepage \appendix\section{The CMS Collaboration \label{app:collab}}\begin{sloppypar}\hyphenpenalty=5000\widowpenalty=500\clubpenalty=5000
\cmsinstitute{Yerevan Physics Institute, Yerevan, Armenia}
{\tolerance=6000
A.~Hayrapetyan, A.~Tumasyan\cmsAuthorMark{1}\cmsorcid{0009-0000-0684-6742}
\par}
\cmsinstitute{Institut f\"{u}r Hochenergiephysik, Vienna, Austria}
{\tolerance=6000
W.~Adam\cmsorcid{0000-0001-9099-4341}, J.W.~Andrejkovic, T.~Bergauer\cmsorcid{0000-0002-5786-0293}, S.~Chatterjee\cmsorcid{0000-0003-2660-0349}, K.~Damanakis\cmsorcid{0000-0001-5389-2872}, M.~Dragicevic\cmsorcid{0000-0003-1967-6783}, A.~Escalante~Del~Valle\cmsorcid{0000-0002-9702-6359}, P.S.~Hussain\cmsorcid{0000-0002-4825-5278}, M.~Jeitler\cmsAuthorMark{2}\cmsorcid{0000-0002-5141-9560}, N.~Krammer\cmsorcid{0000-0002-0548-0985}, D.~Liko\cmsorcid{0000-0002-3380-473X}, I.~Mikulec\cmsorcid{0000-0003-0385-2746}, J.~Schieck\cmsAuthorMark{2}\cmsorcid{0000-0002-1058-8093}, R.~Sch\"{o}fbeck\cmsorcid{0000-0002-2332-8784}, D.~Schwarz\cmsorcid{0000-0002-3821-7331}, M.~Sonawane\cmsorcid{0000-0003-0510-7010}, S.~Templ\cmsorcid{0000-0003-3137-5692}, W.~Waltenberger\cmsorcid{0000-0002-6215-7228}, C.-E.~Wulz\cmsAuthorMark{2}\cmsorcid{0000-0001-9226-5812}
% ...
\par}
\vskip\cmsinstskip
\dag:~Deceased\\
$^{1}$Also at Yerevan State University, Yerevan, Armenia\\
$^{2}$Also at TU Wien, Vienna, Austria\\
% ...
\end{sloppypar}
\end{document}